\documentclass[a4paper,11pt]{article}
\usepackage{pos}
\usepackage{multicol}

\title{The International Lattice Data Grid (ILDG 2.0)}

\author*[a]{Francesco Di Renzo}

\affiliation[a]{DSMFI Università di Parma and I.N.F.N.,\\
  Parco Area delle Scienze 7/A, 43124 Parma, Italy}

\emailAdd{francesco.direnzo@unipr.it}

\abstract{We report on status and perspectives of the International
  Lattice Data Grid. ILDG was established some twenty years ago as a
  community-wide initiative to enable the sharing of gauge
  configurations generated by many major lattice
  collaborations. After a phase in which availability and usage of services had
  degraded, an effort to modernize and reactivate ILDG 2.0 has been started.
  The initiative has made important progress and we can look forward to
  larger and fully FAIR data sets becoming available to a wider audience.}

\FullConference{The 40th International Symposium on Lattice Field Theory (Lattice 2023)\\
July 31st - August 4th, 2023\\
Fermi National Accelerator Laboratory\\}

\begin{document}
\maketitle

\section{Introduction: from ILDG to ILDG 2.0}
\begin{multicols}{2}

The basic data sets from Lattice QCD simulations are gauge
configurations. Their generation takes a large amount of our
computational efforts.
Therefore, the International Lattice Data Grid (ILDG)
was started some twenty years ago as a community-wide 
initiative aiming at making these precious data available to
the international scientific community.

While our keywords in ILDG are to make data ``sharable, usable, and citable'',
it is fair to say that ILDG has been anticipating and implementing
already most of the FAIR principles \cite{FAIR}. They were
formulated in 2016, roughly a decade after ILDG had become operational,
Since then the quest for data to be Findable, Accessible, Interoperable,
Reusable has become a more and more important goal and requirement
in many fields of Science.  

\begin{center}
\includegraphics[width=\linewidth]{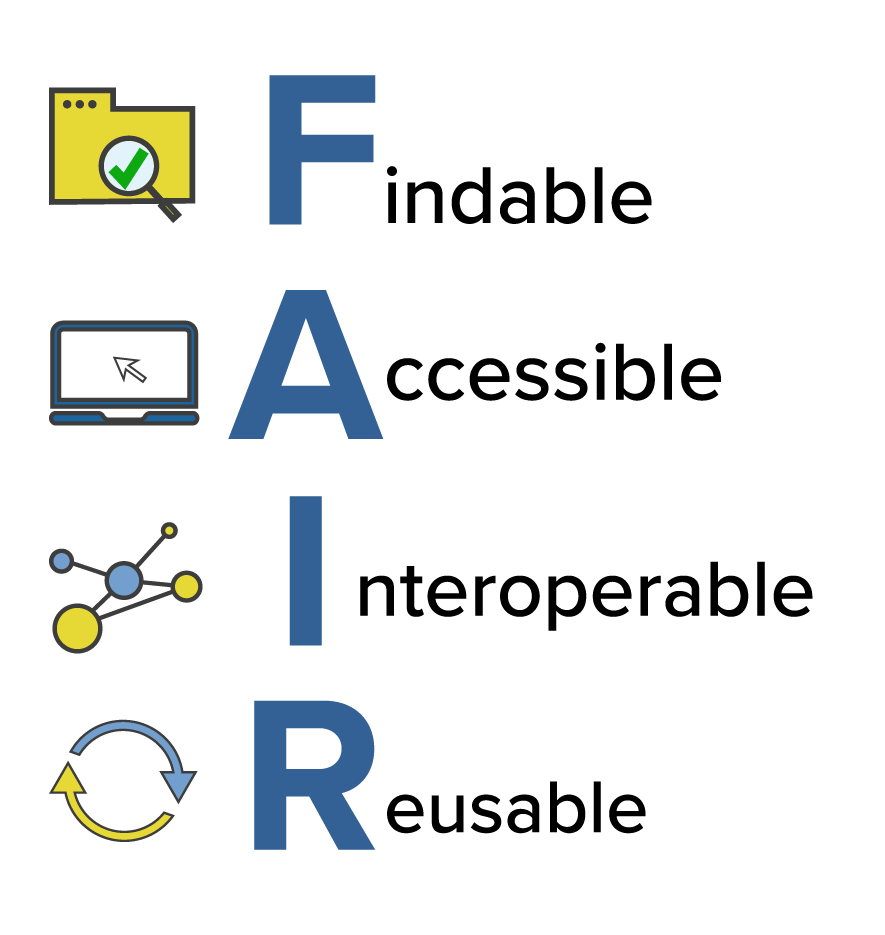}
\captionof{figure}{A popular graphical description of the FAIR acronym.}
\end{center}

\end{multicols}

For Lattice QCD gauge configurations are precious products in 
terms of both human efforts and computing resources. Resources in this
context {\em in primis} means energy and whatever comes with that
(including potentially important contributions to the $CO_2$ budget). 
Apart from making these precious raw data sharable, usable, and citable
for the community, important goals of ILDG are also to promote basic
quality standards for lattice data and to help putting into practice
solid data management plans. 

ILDG is organized as a federation of autonomous {\em Regional Grids},
forming a single {\em Virtual Organisation}. The data sets stored within 
ILDG are ensembles of gauge field configurations, ultimately 
consisting of sets of metadata and binary files; this requires 
community-wide agreed standards. Given a convenient definition of a 
XML schema, metadata are searchable on the web accessing the metadata 
catalogues. Of course, not only metadata are standardized, but also 
data formats.
Moreover, APIs have been developed as well, in particular interfaces 
to data access services. Within the framework thus established, 
distributed data repositories have been over the years deployed and 
are managed by the regional grids. The ILDG being a federation, 
the latter are organized with individual policies and implemented 
with different architectures and technologies. In fact, regional 
grids autonomously operate the basic services: metadata and 
file catalogues, as well as storage elements (and web pages).\\

\begin{figure}[h] 
\centering
\includegraphics[scale=0.5]{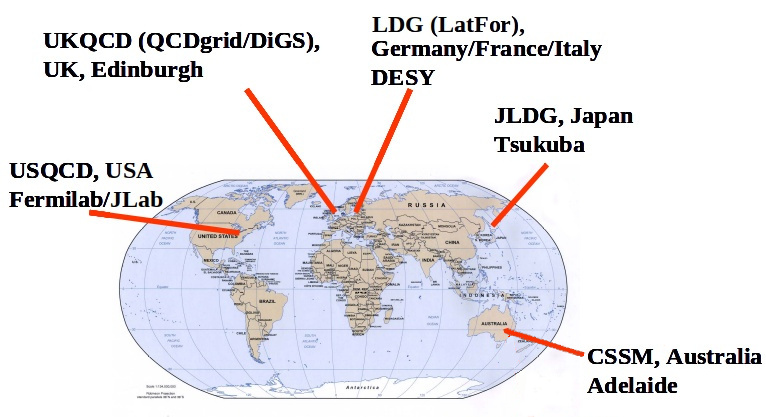}
\caption{The regional grids federated into ILDG.}
\end{figure}

While the usefulness of this effort has always been widely
acknowledged, a few years ago it was recognised that the ILDG 
infrastructure had severely degraded over the years: while 
services were still (at least partially) online, but 
they were usable only by experts. At a time of increasing 
emphasis on Open Science/Open Data paradigms, this was not 
acceptable for a community who used to be at the forefront.
Action was needed and thus a new community effort was started 
to re-activate and modernize the ILDG in terms of organization, 
technology, usability: ILDG 2.0 was thus born.\\
Given the general interest in Open Data issues, it has been 
possible to seize important collaboration opportunities. 
In particular, it was possible to leverage support and funding 
of various national initiatives for ILDG 2.0. A very noticeable 
example is the PUNCH4NFDI initiative in Germany, which ``{\em has 
the objective to systematically index, edit, interconnect and make 
available the valuable stock of data from science and research}'' 
\cite{PUNCH4NFDI}. Its main goal is that of setting up a federated 
science data platform, implementing FAIR principles. This includes
services and interfaces to access and use data and 
computing resources of the many involved scientific communities. 
In Italy, the I.N.F.N. (Istituto Nazionale di 
Fisica Nucleare) has always ``{\em supported open science and 
aimed to raise awareness within its scientific community, promoting 
the adoption of policies in support of open science and the knowledge 
of best practices that support these principles}''
\cite{INFN_OpenData}. Both for 
development of tools and for deployment of resources, I.N.F.N. has
always been at the forefront in this respect and not surprisingly 
is an important partner for ILDG, providing an effective solution
for the future identity and access management.\\

Communications and discussions on the status and perspectives of ILDG
have been regularly included in the program of the annual Lattice conference.
The plans for ILDG 2.0 have been presented in a plenary talk at Lattice
2022, and in \cite{LAT22plenary} the interested reader can find a concise
and quite complete account of basic ILDG concepts, as well as its
compliance with FAIR principles. In the following, we will mostly focus
progress since then.

\section{Current status of ILDG 2.0} 
\subsection{Resuming operation of regional grids}

While ILDG services have never gone (completely) offline, major
progress has been made in resuming their full operation with
the new API. The current status is as follows:
\begin{itemize}
\item {\bf JLDG} (Japan) catalogues are online, and $60$ ensembles
  with about $40K$ configurations from (what we can quote now as) ILDG 1.0
  are available from the Gfarm file system.
\item {\bf LDG} (Continental Europe) catalogues are online, and $259$
  ensembles with about $300K$ configurations from ILDG 1.0 are
  available from several storage elements.
\item at the time of the Lattice 2023 conference, resuming services
  was in progress in the UK and planned in the US.
\end{itemize}

One of the main concept of ILDG has always been that the regional
grids do not necessarily need to, but can use the
same implementation of the {\em metadata and file catalogues}. In view
of this, an interesting component of ILDG 2.0 is a completely re-factored
design of these services; this important development has only become possible
by the work of professional software engineer funded by PUNCH4NFDI.
The new design is fully containerized and thus easy to deploy.

The catalogues implement a REST API according to the new ILDG specifications
and support different configuration options, e.g. of access policies and
additional metadata collections, according to the specific needs the regional
grids or other use cases. A fine-grained control of read and write access,
which so far was only partially supported in ILDG, is a main advantage of the
planned transition to token-based authentication in ILDG 2.0. A flexible
and fine-grained access control mechanism is essential to allow a simple
and smooth switching from collaboration-internal sharing of data, e.g.
during initial embargo periods, to community-wide data access and publishing.

\subsection{Interaction with the lattice community}

A first {\em Hands-on Workshop} was held (online) in June 2023 and
mainly targeted to those collaborations which during the parallel
session on ``Lattice Data'' \cite{LAT22parallel} at the Lattice 2022
conference expressed interest in using ILDG. The workshop was attended
by a remarkable number of participants (37) from 12 collaborations.
In addition to presentations on detailed aspects of the ILDG
metadata and middleware, most participants were able to successfully
carry out hands-on exercises on the basic ILDG functionalities:
search, download, markup, and upload. The presentations of the workshop
are available online \cite{HandsOn1}, and a containerized user
environment together with basic (so far only low-level) client scripts
can be downloaded \cite{HandsOn2}. Some further effort is yet needed
and desirable to extend this material to a self-contained documentation
and tutorial.

While the hands-on workshop provided a very useful and successful first
test of the new middleware setup, an important aim was to to understand
the ILDG-readiness of the various collaboration and to collect information
about features, which currently prevent data providers from uploading new
configurations. The feedback during and after the workshop was of great help
to identify and prioritize issues and missing features, in particular of the
metadata schema and file format. These are now being addressed by the working
groups to prepare a new revision of the corresponding ILDG specifications.

To enable a single point of contact for ILDG users to reach
reach ILDG developers, Working Groups, and Board the email
\href{mailto:ildg-contact@desy.de}{ildg-contact@desy.de} has
been set up. Last but not least, a revision of the ILDG web pages
at \href{https://hpc.desy.de/ildg}{https://hpc.desy.de/ildg}
with basic information on ILDG is in progress. The content still
needs to be improved and extended, and the URL might move in the
future.

\subsection{Organization of ILDG}

The progress of ILDG heavily relies on the participation and contributions
of the user community. The activities are organized and coordinated through
the Metadata and Middleware Working Groups and the ILDG Board, which resumed
frequent and regular meetings.

\begin{itemize}
\item The {\bf Metadata Working Group} is in charge of agreeing on concise and
  community-wide standards for the metadata description and storage of
  the data in ILDG. This requires a careful balancing between
  scientific and technical considerations, and a continuous effort to keep up with
  the constant progress and developments of lattice gauge theory research. The
  resulting specifications and extensions of the QCDml metadata schema (the XML
  schema for marking up gauge configurations is called QCDml) and of the data
  formats, is essential to make ILDG data compliant with the FAIR principles.
\item The {\bf Middleware Working Group} specifies functionality and
  interfaces of the ILDG services in order to guarantee the interoperability
  of the regional grids. Important practical aspects of this work include
  studies of new technologies, the sharing of corresponding expertise, and
  suggestions and prototype implementations of the services to be run by
  the regional grids, as well as of user tools.
\item The activities of two working groups have a high level of synergies
  and interrelation. In fact meetings and discussions of the two groups
  are often done jointly. Therefore, it is fair to mention their members
  (at the time of the Lattice 2023 conference) as a single list:
  {\em T.~Amagasa,
      Basavaraja~B.S., C.~DeTar, B.~Joo, C.~McNeile, O.~Kaczmarek,
      G.~Koutsou, H.~Matsufuru (convenor), Y.~Nakamura, D.~Pleiter, H.~Simma
      (convenor), C.~Urbach, O.~Witzel, T.~Yoshie, J.~Zanotti}.  
\item The {\em Board} represents ILDG towards both the lattice community
  and the service providers. It is in charge of discussing and deciding policies
  and guidelines for membership and data sharing. The Board supports
  regional grids in applying for resources and oversees the activities
  of the working groups. The Board includes representatives of each regional
  grid and current members (at the time of the Lattice 2023
  conference) are: {\em F.~Karsch (chair), Y.~Aoki, B.~Blossier, F.~Di Renzo,
    R.~Edwards, W.~Kamleh, Y.~Kuramashi, D.~Leinweber, A.~Portelli, J.~Simone}.
\end{itemize}

On the way towards ILDG 2.0, the working groups, together with the Board, need
to take care not only of technical, but also of administrative and organizational
matters of ILDG, like the policy framework, the selection of suitable federations
of trusted identity providers, or the protection of personal (user) data. For
instance, both the AUP (Acceptable Use Policy) and the VO-Policy documents had
to be revised to satisfy the latest requirements and feedback form both service
providers and user community. The revision of the VO policy, which was in the
final draft status at the time of the conference, has meanwhile been approved
and put in place \cite{VOpolicy}.

\subsection{Work in progress}

Work on several important tasks, which still need to be accomplished in
order to make ILDG 2.0 fully operational and usable, is currently in
progress:

\begin{itemize}
\item As previously pointed out, the constant progress of lattice
  gauge theory research requires an update of the {\em metadata schema and file format},
  e.g. to allow markup of new actions. A new revision taking into account
  the requirements and feedback from most of the ILDG-ready collaborations
  is expected to be released early in 2024.
\item {\em User-friendly and reliable client tools} are needed in particular
  for complex search operations and easy markup, optionally also with a
  graphical interface. Their development deserves high priority once all
  services are running and finalized, and should include essential feedback
  and contributions from the user community itself.
\item The authentication based on grid certificates is something that
  for sure has contributed to the perception of ILDG as a
  ``not-so-easy-to-access'' environment. While this has probably been
  emphasised more than it would have deserved, the migration from grid
  certificates to {\em token-based authentication} is a
  mandatory step for ILDG 2.0 that is currently being set up and tested
  as illustrated in Fig.~3. The INDIGO IAM
  \cite{IAM} (maintained and developed by INFN/CNAF) has been selected as
  the {\em new Identity and Access Management service}. An IAM instance
  dedicated to ILDG has already been deployed at INFN-CNAF
  (\href{https://iam-ildg.cloud.cnaf.infn.it}{https://iam-ildg.cloud.cnaf.infn.it}).
  We are confident that the new setup will enable a simpler user
  registration and usage of ILDG. However, some further effort is still
  needed to until all relevant home institutions are included in the
  eduGAIN framework \cite{eduGAIN}.
\item It is also important to keep in mind that provisioning of actual storage
  space is responsibility of the regional grids. Efforts in this respect
  need to be continued and supported by collaborations who plan to provide
  their data through ILDG. Moreover, since cloud storage technology is becoming
  increasingly important, studies of how such services can be integrated
  with ILDG are under discussion and need to be further followed up.
\item If a criticism had to be quoted from the very old days, one
  would probably cite a certain perception of ILDG as ``an
  environment for experts''. In view of that, ongoing efforts to
  provide and improve {\em documentation and instructions for non-experts}
  need to be continued and further extended.

\end{itemize}

\begin{figure}[hbt]
	\centering
	\includegraphics[scale=0.2]{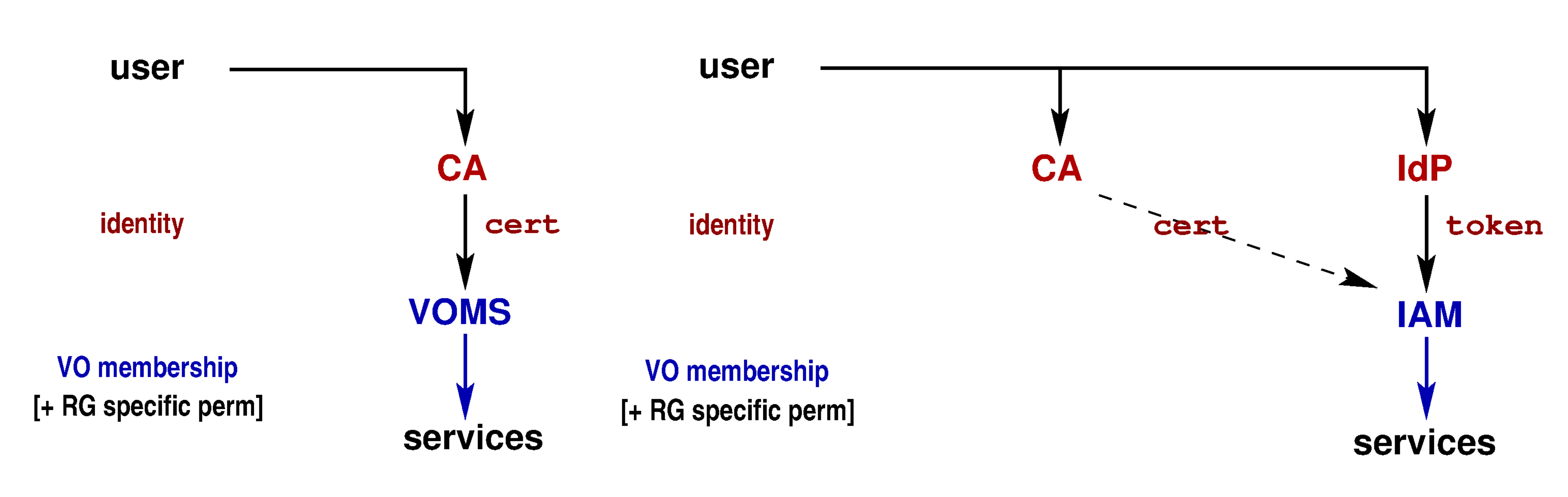}
	\caption{A sketch of the migration from
  authentication based on certificates to a token-based one.}
\end{figure}


\section{Outlook}

Based on the continuous progress in many concurrent efforts on the
re-activation and improvement of ILDG, it is reasonable to look
forward to a renewed, boosted role of ILDG (in its 2.0 version).
This is also reflected and driven by the increasing interest from
many major lattice collaborations to share their gauge configurations
in a FAIR-compliant setup.

From a more ILDG-inside perspective, the intense efforts of the
re-activated working groups and the evidence of more and more ILDG
services coming back into operation can be trusted as evidence that
ILDG 2.0 will become fully up and running in due time. The transition
to simplified and modern access methods and availability of more
documentation can be expected to make ILDG as a useful framework for
community-wide sharing of gauge configurations in Lattice QCD.

Of course, as a an initiative, which is driven and carried out by the
community itself, improving and sustaining ILDG also relies strongly
on the efforts and contributions of the user community itself. To
strengthen this link and provide the required practical training and
know-how, also a further, community-wide Hands-on Workshop is planned.

All in all, there is still (a lot of) work ahead of us, but we can be 
reasonably optimistic. As a realistic goal for the 2024 Lattice conference
we aim at new configurations from at least 5 collaborations being uploaded
and made available through ILDG.

\section*{Acknowledgements}

We first of all thank the Lattice 2023 organisers for having provided
a chance to present the status and prospects of ILDG 2.0. The ILDG is
an initiative which is based on the efforts by so many people that is
impossible to exhaustively mention them all: we acknowledge many valuable contributions by the members of the Metadata and Middleware Working Groups, as well as the ILDG Board. It is a pleasure to explicitly thank Tomoteru Yoshie for his continuous efforts for the ILDG over two decades, the convenors of the Working Group H.~Matsufuru and H.~Simma for bringing momentum into ILDG 2.0, and Basavaraja B.S. for his essential development work for the new catalogue services. We thank Luca Dell'Agnello,
director of CNAF and all the members of the INDIGO IAM team for supporting ILDG. \\
This work was in part supported by DFG fund "NFDI 39/1" for the 
PUNCH4NFDI consortium. We acknowledge funding from the European Union’s
Horizon 2020 research and innovation program under the Marie 
Skłodowska-Curie grant agreement No. 813942 (ITN EuroPLEx).


\begin{thebibliography}{99}
\bibitem{FAIR}
M. D. Wilkinson, M. Dumontier, I. J. Aalbersberg et al., {\em The FAIR 
Guiding Principles for scientific data management and stewardship}, 
Scientific Data. 3 (2016), 160018, doi:10.1038/SDATA.2016.18
\bibitem{PUNCH4NFDI}
\href{https://www.punch4nfdi.de}{https://www.punch4nfdi.de}
\bibitem{INFN_OpenData}
\href{https://web.infn.it/openscience/}{https://web.infn.it/openscience/}
\bibitem{LAT22plenary}
F.~Karsch, H.~Simma and T.~Yoshie,
{\em The International Lattice Data Grid -- towards FAIR data},
PoS \textbf{LATTICE2022} (2023), 244,
doi:10.22323/1.430.0244
\bibitem{LAT22parallel}
G.~Bali, R.~Bignell, A.~Francis, S.~Gottlieb, R.~Gupta, I.~Kanamori,
B.~Kostrzewa, A.~Y.~Kotov, Y.~Kuramashi and R.~Mawhinney et al.,
{\em Lattice gauge ensembles and data management},
PoS \textbf{LATTICE2022} (2022), 203,
doi:10.22323/1.430.0203
\bibitem{HandsOn1}
\href{https://indico.desy.de/event/39311}{https://indico.desy.de/event/39311}
\bibitem{HandsOn2}
\href{https://gitlab.desy.de/ildg/hands-on}{https://gitlab.desy.de/ildg/hands-on}
\bibitem{IAM}
\href{https://github.com/indigo-iam/iam}{https://github.com/indigo-iam/iam}
\bibitem{VOpolicy}
\href{https://hpc.desy.de/ildg/vo_policy}{https://hpc.desy.de/ildg/vo\_policy}
\bibitem{eduGAIN}
\href{https://edugain.org}{https://edugain.org}

\end{thebibliography}
\end{document}